\documentclass[10pt]{article}
\usepackage{geometry}                 
\usepackage{graphicx,multicol}
\usepackage[italian]{babel}

\usepackage{amssymb,bm,hyperref,color}
\usepackage{cite}
\usepackage{nicefrac}
\usepackage{bbold}

\title{Spunti dalla storia per insegnare l'anti-materia}
\author{Francesco Vissani \\
\small INFN, Laboratori Nazionali del Gran Sasso}
\date{}                                           
\begin{document}
\maketitle

\begin{abstract}
Il concetto di antimateria \`e importantissimo, ma non sempre 
discusso come meriterebbe, bilanciando idee e formalismo. In questa nota raccogliamo alcuni spunti per presentarlo in modo efficace, che ricalcano certi passi effettuati nella storia della conoscenza; sebbene ricordati solo di rado, possono servire per arricchire  il materiale 
didattico standard. Oltre agli arcinoti contributi di Dirac, che collochiamo nel loro contesto originale, risaltano quelli di Pauli e soprattutto di Majorana, che per primo pervenne al moderno formalismo della quantizzazione canonica. Emerge l'importanza del punto di vista della meccanica ondulatoria, che pure mostra i suoi limiti, richiedendo qualche modifica per costituire una interpretazione accettabile.

\end{abstract}

%
\parskip0.7ex
 
\section{Introduzione}
 
Mentre non \`e raro che si parli di anti-materia, \`e meno comune che questa nozione venga discussa a fondo. 
Quando ne ragioniamo col grande pubblico diamo atto delle osservazioni che ne comprovano 
l'esistenza e magari parliamo della applicazioni in medicina, senza azzardarci a  spiegarne l'origine.
Se ne parliamo in un corso di fisica delle particelle, lo consideriamo ovvio,  presentando l'ampia ``zoologia'' 
senza soffermarci  troppo a ragionare di un aspetto tanto generale.
Nei corsi 
universitari di natura teorica, infine,  introduciamo l'anti-materia solo in coda ad un faticoso percorso intellettuale,  e di regola adottiamo  approcci
apparentemente  privi di punti di contatto coi  corsi precedenti, rendendo oneroso se non  impossibile districare  idee e schemi matematici.
Lo scopo di questa nota \`e quello di  presentare con chiarezza tale concetto, tratteggiando un percorso didattico di livello universitario basato 
sulla storia della conoscenza, che ricorre  per quanto possibile ai {\em formalismi e linguaggi} 
della meccanica quantistica non relativistica gi\`a noti allo studente, per giungere gradualmente fino a quelli tipici della meccanica quantistica relativistica.
 
\subsection{Assunzioni sulle conoscenze gi\`a acquisite e terminologia}
Dietro il termine ``anti-materia'' si nasconde una lunga e complessa discussione.  Come vedremo, per 
fare completa chiarezza, bisogna distinguere tra la nozione teorica di  ``anti-elettrone'' e quella sperimentale di ``positrone''; e vedere come 
i termini originariamente introdotti 
vengono poi riutilizzati per riferirsi a concetti, schemi  e interpretazioni 
 {\em nuovi e diversi}. 
 
Ma continuiamo ad esaminare attentamente la terminologia corrente. 
Grazie ad un formalismo  consolidato dalla pratica e dalle discussioni, oggi di solito si ragiona 
di particelle ed antiparticelle, senza fare alcun riferimento allo spin. Ci si riferisce dunque sia ai fermioni che ai bosoni; basti pensare ai pioni carichi $\pi^\pm$ o anche 
bosoni $W^\pm$, che sono indiscutibilmente coppie di particelle ed anti-particelle,
pur non essendo materia o anti-materia.

In  gran parte di questo lavoro non ci  interesseremo del generale concetto di anti-particella (a cui si pervenne nel corso degli anni) 
ma di quelli,  pi\`u specifici e tuttavia molto importanti,  di anti-elettrone,   di anti-materia, di particelle che costituiscono l'anti-materia. 
Sono nozioni  che prendono forma nel corso dello sviluppo della teoria dell'elettrone atomico, e precisamente 
durante la discussione dell'equazione d'onda a cui obbedisce l'elettrone relativistico, l'equazione di Dirac.

Per apprezzare fino in fondo tutti questi concetti, credo sia utile capire quali ostacoli i fisici 
dovettero affrontare e superare, mentre procedevano 
nella comprensione delle prime particelle che incontrarono: gli elettroni. 
All'inizio si sperava che  essi fossero una sorta di microscopici baluardi dell'esistenza; 
questa aspettativa \`e esplicita in Thomson, che  li denomina sempre  ``corpuscoli''. 
Ma presto iniziano ad emergere alcune strane caratteristiche dell'elettrone, tra cui lo spin ed  il carattere statistico 
riassunto col termine ``fermione''; infine, all'inizio degli anni trenta, i fisici si trovano a dover accettare    
 una sua inquietante controparte, che evoca i {\em doppelg\"anger} della letteratura.
 
Ricordiamo che era stato  Pauli il primo  a ipotizzare un numero quantico a due valori (1924), 
senza corrispondenti nella fisica nota, che gli consent\`i di formulare  subito dopo 
il principio di esclusione per gli elettroni atomici. L'anno seguente,  tale numero quantico ricevette, grazie a Goudsmit e Uhlenbeck, 
l'interpretazione in termini di spin. Nel 1927, Pauli estese l'hamiltoniana di Schr\"odinger includendo 
le interazioni tra spin e campo magnetico descrivendo cos\`i i fenomeni osservati in fisica atomica.  Ma 
lo spin dell'elettrone continuava ad essere percepito come un carattere 
sfuggente o misterioso, cos\`i come la connessione col suo carattere statistico. 

Sono argomenti esposti nei corsi universitari di fisica atomica e/o di meccanica  quantistica.  
Immaginiamo dunque uno studente universitario che abbia gi\`a acquisito tali concetti e che 
si appresti ad affrontare, per esempio, un corso di fisica teorica delle particelle. Volendo preparare il terreno, 
si potrebbero  evidenziare sin dall'inizio due aspetti  importanti aspetti 
che, generalizzando le caratteristiche del sistema elettrone - positrone, caratterizzano la nozione  moderna di  anti-materia:
\begin{itemize}
\item  ci si riferisce all'esistenza di una  
coppia di particelle con 
 masse  uguali e cariche elettriche opposte;
\item entrambe le  particelle hanno spin $\nicefrac{1}{2}$ e soddisfano la statistica 
di Fermi-Dirac.
\end{itemize}
Il primo aspetto si riferisce a particelle ed  anti-particelle in genere,  e pu\`o essere ancora generalizzato considerando, invece della carica elettrica, 
una carica diversa - p.e., la carica barionica o leptonica. 
Il secondo precisa la definizione delle particelle di materia e di
anti-materia, sulle quali ci  focalizzeremo. Questi caratteri, inizialmente riscontrati per l'elettrone (e la sua controparte), risultano
tra loro associati per ogni particella carica; vedremo come i vari schemi teorici tengono conto o descrivono questo fatto.
 Il caso dei fermioni neutri, particelle di materia prive di carica elettrica (come il neutrone ed il neutrino)  merita 
una discussione speciale, che svolgeremo   dopo aver presentato il modo in cui i vari schemi trattano 
l'elettrone/positrone ed in generale le particelle cariche.

Discuteremo come queste nozioni presero forma e come,  
per mezzo di successive elaborazioni, si arriv\`o a comprendere  
in modo soddisfacente l'antimateria.

\subsection{Un risultato ed un problema di Dirac}
Nel 1928, Dirac esib\`i la celeberrima equazione d'onda che descrive 
un elettrone libero, che obbedisce ai principi della relativit\`a di Einstein \cite{d28}.
Egli mostr\`o che tale equazione 
\begin{itemize}
\item include in modo automatico lo spin dell'elettrone, e 
\item predice 
il corretto accoppiamento con il campo magnetico  (rapporto giro-ma\-gne\-tico).\footnote{La possibilit\`a di descrivere il protone ed il neutrone 
con la stessa equazione d'onda  rest\`o per qualche tempo in dubbio a causa dei valori osservati dei loro rapporti giro-magnetici diversi da quelli previsti usando l'equazione di Dirac 
(``anomali'').}
\end{itemize}
Questi risultati   vennero considerati come indiscutibili progressi alla discussione; 
essi si presentavano come una sorta di  coronamento delle indagini della fisica atomica.

Tuttavia Dirac era ben cosciente che la sua equazione differisce da 
quella di Schr\"odinger per un cruciale aspetto: in apparenza, 
non esiste uno stato di minima energia. Infatti, 
l'intero spettro degli stati liberi con energie positive risulta essere duplicato in quelle negative. 
Tali circostanze, 
prese letteralmente,  sembrerebbero implicare che gli atomi non esistono: un elettrone atomico  a prima vista potrebbe 
accedere a stati con energie sempre pi\`u basse, un po' come ci si aspetterebbe avvenga col 
modello classico dell'atomo planetario. 

Esamineremo  la via d'uscita proposta da Dirac 
confrontandola con altre successive, che lo porta nel 1931 ad introdurre per la prima volta 
il termine ``anti-elettrone'' \cite{d31}. 
Ribadiamo per massima chiarezza che la particella che   venne presa a modello 
 per  sviluppare la discussione  teorica dell'anti-materia \`e l'elettrone,  e tale discussione 
ricevette una vertiginosa accelerazione dalla  scoperta del positrone, avvenuta 
appena un anno dopo  
la proposta avanzata da Dirac \cite{a32}.

\subsection{Un argomento successivo}\label{alt}
Naturalmente, non \`e necessario dover ripercorrere per intero il faticoso cammino della storia. 
Decidendo di procedere cos\`i, per non subire la suggestione dei precedenti argomenti (ovvero per coltivare  vantaggiosamente il seme del dubbio)  
si pu\`o partire proponendo anche  
un altro ragionamento, molto semplice, ma  che 
emerge dopo un lungo dibattito teorico \cite{pw34,m37,p41,s41,s42,f49}.
Esso 
mette in chiaro uno dei due aspetti cruciali
dell'idea corrente di anti-materia e 
pu\`o essere presentato prima di introdurre  un formalismo pi\`u  completo e soddisfacente.
 
Si immagini che ci sia un nucleo atomico, o magari un  semplice neutrone, soggetto all'emissione di elettroni (decadimento $\beta$)\footnote{Questa possibilit\`a, alla quale ogni moderno fisico \`e assuefatto, cela un passaggio concettuale piuttosto drammatico: l'idea che particelle di materia possano essere create o distrutte. Dal punto di vista storico, un importante passo avanti verso la sua accettazione 
\`e costituito dalla teoria di Fermi del decadimento $\beta$ \cite{f33} che ricevette supporto intellettuale dall'idea di de Broglie 
che le particelle di materia possano essere pensate come onde.} e consideriamo
per amore di discussione il caso, suggerito dall'equazione di Dirac, 
che l'elettrone abbia energia negativa. Il neutrone  diventer\`a un protone, conservando cos\`i la carica elettrica; inoltre la sua energia {\em aumenter\`a}, in quanto stiamo considerando l'emissione di una particella con energia negativa. L'osservazione cruciale \`e che gli  
stessi identici effetti seguono 
se, invece di emettere questo ipotetico elettrone,  il neutrone {\em assorbe} un positrone,  diventando in ogni caso un protone. In effetti, questo accresce
tanto la carica del nucleone quanto la sua energia.

Questa  considerazione suggerisce la possibilit\`a di operare una sistematica e conveniente sostituzione di concetti: anzich\'e pensare ad  un elettrone 
di energia negativa  emesso assieme al protone,  possiamo  
pensare ad una particella con carica positiva, inizialmente  
coesistente col neutrone, e che venga poi da esso assorbita. 
Da questo punto di vista, non sembra 
inevitabile parlare di particelle con energia negativa, come l'equazione di Dirac sembrerebbe pretendere. Per\`o  ci si trova 
obbligati ad ammettere che a fronte di ogni particella carica ce ne debba essere sempre un'altra con carica opposta. 
Per ulteriori discussioni, referenze, e grafici illustrativi rimando a  \cite{libro,majo}.

\subsection{Organizzazione di questa nota}
Assumendo che l'equazione di Dirac sia stata gi\`a esposta, come  avviene nei corsi universitari di fisica, 
esplicitiamo  alcune osservazioni supplementari di carattere matematico di particolare rilevanza per trattare i fermioni relativistici. 

Iniziamo descrivendo un modo molto diretto   per introdurre  l'anti-materia, che riprende le idee del paragrafo precedente e  si basa  
su una specifica modifica del formalismo della meccanica quantistica ondulatoria (sez.~\ref{s-ond}).
Procediamo  esponendo quello che fu effettivamente usato da Dirac (1931)  e  che  - dopo la scoperta di Anderson - venne 
utilizzato dall'intera comunit\`a scientifica per diversi anni (sez.~\ref{s-dir}).
Mostriamo infine come la proposta avanzata da Majorana nel 1937 
non si limiti alla sostituzione di concetti descritta appena sopra, ma 
renda molto pi\`u armonioso e coerente il quadro teorico, spiegando anche il carattere fermionico di elettroni e positroni
 (sez.~\ref{s-maj}).

\begin{table}[t]
\centerline{\small
\begin{tabular}{|ll|c|cc|c|}
\hline
autori & data & nome &   spin  & principio di  &  menzionato  \\
 & pubbl. & usato  &$\nicefrac{1}{2}$ & esclusione  &   in  \\ \hline
Pauli-Weisskopf & 1934 &  - & no & no  & ( sez.~\ref{s-ond} ) \\
Stueckelberg & 1941 & anti-particella  & ignorato &   ignorato  & ( sez.~\ref{s-ond} )\\
Dirac  &  1931 & anti-elettrone   &  necessario & imposto  & sez.~\ref{s-dir}  \\ 
Majorana &  1937 &  positrone & necessario & necessario & sez.~\ref{s-maj} \\ \hline
\end{tabular}}
\caption{\em\small Schemi di quantizzazione relativistici.
I primi due, che menzioniamo per completezza 
anche se {\em non trattano gli elettroni,} sono quello di Pauli che 
 descrive degli scalari (e da lui detto teoria ``anti-Dirac'') e quello di Stueckelberg che si interessa al concetto pi\`u generale di anti-particella.
Gli altri, 
basati sull'equazione di Dirac e discussi in dettaglio in questa nota, spiegano lo spin dell'elettrone.
Dirac {\em invoca}  il principio di esclusione solo in un secondo tempo, ed in questo senso non lo spiega.
La procedura di Majorana include in modo automatico il principio di esclusione; si noti che egli, con grande correttezza,  
usa il termine positrone invece di anti-elettrone,  che {\em identificava} il discutibile costrutto teorico proposto da Dirac.
\label{trab}}
\end{table}

Tale costruzione 
corrisponde  in senso stretto alla moderna  quantizzazione canonica dei fermioni, 
e fornisce il primo quadro concettuale nel quale si descrive in modo del tutto soddisfacente l'anti-materia. 
Non di rado, essa \`e l'unica che viene presentata nei corsi introduttivi universitari, ma includendo
certe sofisticazioni  proposte da Pauli nel 1941 (di cui parleremo)
che mi sembra  
amplifichino le difficolt\`a didattiche invece di attenuarle. Al contrario,   
nella presente nota ci prefiggiamo  di ridurre al minimo le ostruzioni formali,  seguendo da vicino gli argomenti originari, tranne   
minimi  adattamenti per aderire alle  pi\`u comuni convenzioni  vigenti.

La tabella~\ref{trab} riassume lo schema dell'esposizione, anticipando i punti che verranno discussi.

Concludiamo 
con una nota sul caso delle particelle 
senza carica elettrica (sez.~\ref{s-sce}),
uno specchietto storico riassuntivo (sez.~\ref{s-sto}) e 
dei commenti 
sulla presente proposta didattica (sez.~\ref{s-dis}).
Le appendici offrono materiale per familiarizzare: 
$i.$ con la letteratura secondaria   (App.~\ref{appb});
$ii.$ con le matrici nella rappresentazione di Majorana 
(App.~\ref{appa}).

Prima di procedere, segnalo alcune introduzioni
ben leggibili sugli aspetti che tratteremo, 
tra cui il libro \cite{kragh}, 
il lavoro di rassegna \cite{g93}  e 
il capitolo \cite{mj}.
 Ricordo  poi che nel centenario della nascita della moderna meccanica quantistica (2025)  
\`e stato proclamato il ``Quantum Year''  \cite{ab}
e la SIF ha messo a disposizione i pi\`u rilevanti articoli sull'argomento 
pubblicati sul Nuovo Cimento \cite{sif},
inclusi i due cruciali  lavori di
Fermi \cite{f33} e soprattutto di Majorana \cite{m37}, sui quali torneremo a  pi\`u riprese nel seguito.

\section{Equazione d'onda dell'elettrone relativistico e anti-materia}\label{s-ond}
Per iniziare, esponiamo un modo moderno di concepire l'anti-materia,  che 
emerse pian piano nel corso della discussione fino ad imporsi all'attenzione generale \cite{pw34,m37,p41,s41,s42,f49}, 
 in quanto \`e quello pi\`u accessibile dopo un corso di meccanica quantistica non relativistica. 
Esso si appoggia ai concetti della meccanica ondulatoria, integrandoli con le considerazioni esposte nella sez.~\ref{alt}, opportunamente formalizzate.
Questa procedura minimale evidenzia le similitudini e le differenze con le corrispondenti situazioni che si incontrano nella 
meccanica  quantistica non relativistica.

\subsection{Il punto di vista della teoria ondulatoria}

Consideriamo lo schema di Schr\"odinger, nel quale le funzioni d'onda $\psi$ evolvono nel tempo. 
L'hamiltoniana di Dirac $E$, che descrive la propagazione di un elettrone, 
o in generale  di una particella libera con 
 massa $m$,\footnote{Includere il termine di massa \`e essenziale per la fisica atomica. Tuttavia,  
 nella speranza di procedere nella discussione della teoria relativistica (e partecipando alla discussione critica del risultato di Dirac) 
 Weyl studia 
sin dal 1929 il caso in cui lo spin \`e incluso ma la massa \`e zero \cite{w29}. In tal modo si imbatte 
nella  chiralit\`a e nella sua connessione con l'elicit\`a, punti che  avranno feconde applicazioni tre decenni dopo.}  pu\`o essere scritta come: 
\begin{equation}
E\, \psi=\Delta \psi \mbox{ con } \Delta=  c\vec{p} \ \vec{\alpha}+ m c^2 \beta   \label{ed}
 \end{equation}
e dove che $ (\Delta { \psi} )_a=\Delta_{ab} { \psi}_b$; la somma sull'indice spinoriale $b=1,2,3,4$ \`e sottintesa. Porremo:
\begin{equation}
E= i\hbar \frac{d}{dt} \mbox{ e }  \vec{p}= -i\hbar \vec{\nabla} = -i\hbar \left( \frac{d}{dx}, \frac{d}{dy}, \frac{d}{dz} \right)
\end{equation}  
Senza ripetere le motivazioni a favore dell'equazione di Dirac, che sono gi\`a bene 
esposte nei corsi universitari, vorremmo limitarci ad alcune utili annotazioni aggiuntive,
ragionando  un po' pi\`u a fondo sull'operatore differenziale lineare appena introdotto.

Come \`e ben noto, e come venne immediatamente mostrato da Dirac \cite{d28}, 
le 4 matrici   hermitiane $\vec{\alpha}$ e $\beta$ 
realizzano lo spin in un modo compatibile con gli esperimenti e con la relativit\`a speciale.
Ma aggiungiamo seguendo Majorana  una importante osservazione. Anche se 
tali matrici possono essere scelte in molti modi,   ci sono diversi vantaggi a far 
s\`i che le prime tre siano reali e la quarta immaginaria pura, o, equivalentemente, che
le prime tre siano simmetriche e l'ultima asimmetrica:
\begin{equation}
\vec{\alpha}^*=\vec{\alpha} \  ,\  \beta^*=-\beta 
\quad \Leftrightarrow \quad
\vec{\alpha}^t=\vec{\alpha} \  ,\  \beta^t=-\beta \label{pm}
\end{equation}
Per una dimostrazione esplicita di questa possibilit\`a 
si veda l'appendice~\ref{appa} di questo lavoro.
Questa scelta di matrici mostra che l'espressione data in~eq.~\ref{ed}
 equivale ad una equazione differenziale {\em reale;} un fatto che evidenzia l'analogia formale con le equazioni di Maxwell.
 
Consideriamo  le cosiddette ``soluzioni ad energia negativa'', 
che dipendono dal tempo come segue
\begin{equation}
\varphi_-(\vec{x})\ e^{+ i \varepsilon_p t/\hbar} \label{bot}
\end{equation}
Nel nostro caso, l'espressione dell'autovalore dell'energia $\varepsilon$ 
in funzione del momento  $p$ \`e 
\begin{equation}
\varepsilon_p=\sqrt{(cp)^2+(m c^2)^2}\label{essoeps}
\end{equation}
Il pedice della funzione $\varphi_-(\vec{x})$ ci ricorda il segno dell'energia.
 La scelta di matrici in eq.~\ref{pm}  
rende evidente la possibilit\`a di porre la seguente definizione 
\begin{equation}
(\psi_-)^* \stackrel{!}{=} \varphi_-(\vec{x})\ e^{+  i\varepsilon_p t/\hbar}  \mbox{ cosicch\'e }
\psi_-\sim e^{-  i\varepsilon_p t/\hbar} \label{top}
\end{equation}
che, detto a parole, corrisponde a concepire tali soluzioni come 
{\em coniugate} di soluzioni ad energia positiva.
(Subito sotto chiariremo 
il significato del pedice apposto alla funzione~$\psi_-$.)
Tale punto di vista viene evidenziato da Stueckelberg nel 1941 \cite{s41,s42}, essendo implicito nel lavoro di Majorana del 1937 \cite{m37}
ed in quello di Pauli e Weisskopf del 1934 \cite{pw34}. 
Ricordiamo che nel formalismo della meccanica quantistica la funzione d'onda e la sua coniugata corrispondono ai vettori ``ket'' e ``bra'' 
del formalismo di Dirac, e possono essere pensate rispettivamente come uno stato iniziale e finale di una transizione. 
Vedi \cite{what21} per una pi\`u articolata presentazione del punto.

Lungo tutto il seguente testo, adotteremo la scelta di matrici descritta dalle eq.~\ref{pm}. 
L'interesse in questa possibilit\`a  
fu  sottolineato con convinzione da Ettore Majorana~\cite{m37}.
Pauli, che pure aveva effettuato un precedente ed esaustivo studio matematico
delle matrici di Dirac~\cite{p36-1}, riconosce  
il pregio di tale scelta nel lavoro con Fierz~\cite{p39}.

\subsection{Carica elettrica della funzioni d'onda coniugata}
L'equazione differenziale che include le interazioni elettromagnetiche \`e\footnote{La procedura per accoppiare il campo elettromagnetico 
 si basa sulle idee di Weyl, Fock ed altri \cite{oki} e 
viene detta  ``rimpiazzamento minimale'', in quanto comporta la ridefinizione del quadri-momento
$P=(E/c,\vec{p})$ in $P -q A/c $.  
Si noti che questa procedura non \`e invariante sotto coniugazione in quanto
 $(P)^*=-P $ mentre $qA^*=qA$.}
 \begin{equation}
E\, \psi=\Delta_q \psi\mbox{ dove }
\Delta_q= \left(c\vec{p}- q  \vec{A}  \right) \vec{\alpha}+ m c^2 \beta + q  \varphi 
 \end{equation}
Notiamo che 
\begin{equation}\label{pazz}
(\Delta_{+q})^* = -  \Delta_{-q} 
 \end{equation}
 in quanto il quadri-potenziale elettromagnetico $A=(\varphi,\vec{A})$ \`e 
costituito da funzioni reali e la carica elettrica $q$ \`e pure reale, mentre sia $E$ che $\vec{p}$ sono immaginari puri.
Pertanto, quando si accetta l'identificazione delle onde coniugate descritta in eq.~\ref{top},
occorre scambiare il segno della carica $q$. 
 In questo modo, 
mentre continueremo a concepire le funzioni d'onda ``con energia positiva'' come stati iniziali di una reazione, 
re-interpreteremo le funzioni d'onda ``con energia negativa'' come stati finali  con carica opposta della stessa reazione,
grazie alla identificazione  discussa in eq.~\ref{top}.
Avremo dunque due tipi di particelle con cariche opposte che distinguiamo con  i simboli $\psi_+$
e $\psi_-$: 
\begin{equation}
E\, \psi_+=\Delta_{+q} \psi_+
\mbox{ e }
 E\, \psi_- =\Delta_{-q} \psi_-
\end{equation}
Con questo  formalismo minimale possiamo iniziare a parlare di anti-materia nel contesto della teoria ondulatoria,
la stessa a norma utilizzata nei contesti non-relativistici. Esso ci consente di apprezzare che 
la particella di antimateria ha esattamente la stessa massa, ed anche carica uguale e opposta a quella della particella 
di partenza. 

C'\`e tuttavia almeno una  specifica riserva  della quale  si dovr\`a dare atto:
in questo tipo di presentazione il carattere fermionico della particelle a spin  $\nicefrac{1}{2}$ 
resta in secondo piano, e richiede una ipotesi aggiuntiva, arrivando persino a sembrare  estraneo all'equazione di Dirac. 
In effetti, il  tipo di procedura qui descritto corrisponde ai risultati 
ottenuti da Pauli e Weisskopf nel caso  formalmente impeccabile  ma ipotetico 
di una particella con spin nullo \cite{pw34}, che non ci aiuta a parlare di elettroni.
Nel senso sopra chiarito, 
\`e corretto dire che questa procedura ci consenta di parlare di anti-particelle, ma non esattamente di anti-{\em materia}. In termini pi\`u tecnici, questo formalismo non 
equivale (\`e in un certo senso inferiore) alla procedura di quantizzazione dei fermioni, siccome non  giustifica  quale sia il carattere statistico delle particelle
di materia. 
Apprezziamo cos\`i sia i punti di forza che i limiti della trattazione delle anti-particelle basata sulla funzione d'onda.

\section{Mare di Dirac e  quantizzazione dei campi di materia}\label{s-dir}
Sebbene la presentazione della precedente sezione,  
che  segue \cite{what21} e prende vantaggio del punto di vista ondulatorio,
sia il modo pi\`u graduale e forse anche pi\`u efficace per introdurre l'idea di anti-materia in un corso universitario, essa 
lascia qualcosa a desiderare; e inoltre  (\`e bene ripeterlo) 
non corrisponde affatto al modo in cui procedette Dirac. Lo
rievochiamo qui di seguito per il suo significato  storico, epistemologico e   didattico.
Vedremo che il carattere fermionico dell'elettrone assume una grandissima importanza.

\subsection{La teoria dei buchi}
Considerando  la presenza  di stati di energie negative nello spettro della sua equazione d'onda, 
e forse ispirato dalla teoria degli elettroni nei metalli  
che in quegli anni  stava vedendo la luce, 
Dirac si trov\`o a postulare che tutti gli stati di energia negativa fossero gi\`a occupati. 
 Insomma, per inibire l'accesso a tali stati
propose una ipotesi supplementare, ispirata dal principio di esclusione di Pauli. Questa proposta,  chiamata {\em  mare di Dirac} e a prima vista alquanto inquietante - 
ognuno di noi si troverebbe inconsapevolmente immerso in un mare infinito di elettroni - 
era  tuttavia compatibile con l'idea che un nuovo processo potesse avvenire. Fornendo sufficiente energia  $\varepsilon_p+\varepsilon_q>0$ ad 
un elettrone del mare con energia $-\varepsilon_p \le -m_e c^2<0$, esso avrebbe potuto acquistare energia $+\varepsilon_q\ge m_e c^2>0$. Inoltre si sarebbe formata una lacuna nel mare, che si sarebbe comportata come una particella di carica opposta a quella dell'elettrone e  
con energia $\varepsilon_p>0$; 
per l'appunto, la nozione di ``anti-elettrone'' come originariamente proposta. Si trattava della cosiddetta {\em teoria dei buchi};\footnote{In precedenza, Dirac aveva aveva  esplorato l'idea 
che i buchi potessero svolgere il ruolo dei protoni~\cite{d30}, nutrendo insomma la speranza di poter spiegare  
l'esistenza di tutte le particelle di materia note all'epoca con una sola ``sostanza''. Ma Oppenheimer (1930) gli fece notare
che in quel caso l'atomo di idrogeno non sarebbe stato stabile, e Weyl (1931) argoment\`o 
che la massa della particella corrispondente al buco doveva essere uguale a quella dell'elettrone. Dirac accolse le obiezioni e nel 1931 decise di 
proporre su queste basi il costrutto teorico di anti-elettrone.} 
una teoria del 1931 che inizialmente sembr\`o  corroborata dalla scoperta di Anderson del ``positrone'' (1932), al punto che  venne celebrata 
dal Nobel in fisica del 1933.

\subsection{Seconda quantizzazione dei fermioni relativistici}\label{vedq}
Lungo le linee concettuali sopra tratteggiate 
 si giunse alla proposta di un primo tipo di campi quantizzati di cui parliamo qui di seguito, 
che si appoggiava al formalismo  per descrivere 
la transizioni tra stati di singola particella\footnote{Non lo richiamo in  dettaglio in queste note, in quanto viene esposto 
nei percorsi universitari, tipicamente presentandolo con la dizione `spazi di Fock'.} 
sviluppato da Jordan, Wigner, Klein, Heisenberg e Fock \cite{jk,jw,hei31,fo32}.

La versione relativistica  
venne presentata    da Enrico Fermi  ed applicata  in modo 
assai innovativo nel lavoro sul decadimento $\beta$ \cite{f33}. Il campo quantizzato
 \`e dato  dalla somma
\begin{equation}
{\bm \psi}=\sum_s  {\bm a}_s \  \psi_{s}  \label{sq}
\end{equation}
dove 
$s$ sono tutti i possibili stati dell'elettroni e 
${\bm a}_s$ sono gli operatori  che diminuiscono il numero di particelle (operatori di annichilazione).
Si noti che le funzioni d'onda possono essere normalizzate {\em \`a la Born} come nel 
caso non-relativistico, $\int \psi^\dagger \psi\, d^3x=0$, mentre gli operatori, normalizzati dalla condizione $\{ \bm{a}_s,\bm{a}^\dagger_s\}=1$,   producono 
stati normalizzati da quello di vuoto (Fock).

L'operatore ${\bm \psi}$ \`e costruito usando tutte le autofunzioni del  dato operatore differenziale.
Esso  \`e una sorta di {\em catalogo} di tutti i possibili stati dell'elettrone
ai quali si possa accedere; tanto  quelli di energia positiva quanto  quelli di energia negativa. 
Fermi 
li include (interpretando l'hamiltoniana di Dirac in modo letterale) 
 ed utilizza l'ipotesi del mare di Dirac per dar loro senso \cite{quad}.  
Nel lavoro del 1933 si riferisce a questa procedura con la locuzione:  
{\em metodo di Dirac-Jordan-Klein  delle ampiezze di probabilit\`a quantizzate,} 
mentre nei successivi lavori 
 cita \cite{jk}  e \cite{hei31} e parla di: 
{\em metodo di Dirac-Jordan-Klein detto della \guillemotleft seconda quantizzazione\guillemotright.}
Sembra preferibile la seconda denominazione, che si riferisce all'operatore costruito dall'insieme di funzioni d'onda,
denominazione   ancora in uso in fisica nucleare ed altri contesti non relativistici.

In presenza di stati con energie negative, esisteranno ampiezze  di transizione del tipo
 \begin{equation}
 \langle \mbox{mare}\; \mbox{-}s | {\bm a}_s | \mbox{mare}\rangle = 1
 \end{equation}
dove il vettore di tipo ``ket'' 
$| \mbox{mare}\rangle$ \`e lo stato di vuoto che descrive il mare di Dirac, 
$s$ \`e lo specifico stato con energia negativa che viene liberato,  
e dove lo stato finale, il vettore a sinistra di tipo ``bra'' $ \langle \mbox{mare} \; \mbox{-}s|$, indica l'avvenuta formazione di una lacuna nello stato $s$
 del mare.
Concependo cos\`i 
l'anti-elettrone~$\beta^+$, 
Wick \cite{wi34} previde  l'esistenza di 
reazioni come  ${}^{38}_{19}$K$\to {}^{38}_{18}$Ar$+\nu_e+\beta^+$.

\subsection{Dettagli formali e commenti}\label{squ}

Possiamo precisare la posizione esposta in eq.~\ref{sq}
come segue
\begin{equation}
{\bm \psi}(x)=\sum_s  {\bm a}_s(t) \ \psi_{s}(\vec{x})
\end{equation}
Dunque ci si pone nello schema di Heisenberg, nel quale gli operatori ${\bm O}$
evolvono nel tempo e le funzioni di stato restano costanti.\footnote{Quando si includono le interazioni diventa conveniente adottare  
lo schema di interazione, nel quale gli operatori
evolvono secondo l'hamiltoniana libera.} 
Naturalmente, invece di 
postulare l'equazione di Schr\"odinger per gli stati $s$, 
$i\hbar\, \dot{|s(t)\rangle} ={\bm H}|s(t)\rangle$,  ora postuleremo quella di Heisenberg
\begin{equation}
i\hbar\, \dot{{\bm O}}(t) =[\,  {\bm O}(t), {\bm H}\,]
\end{equation} 
In questo modo,  la dipendenza dal tempo 
dei valori di aspettazione  $O(t)=\langle s| {\bm O}|s\rangle$ resta  immutata,
in quanto, nel caso di interesse, l'hamiltoniana ${\bm H}$ non dipende dal tempo.

Siccome le funzioni d'onda $\psi_{s}$ formano un sistema di stati ortonormali, 
si possono ottenere gli ``oscillatori'' ${\bm a}_s$ con energie positive o negative 
 a partire dai campi di seconda quantizzazione per mezzo di una sorta di proiezione o prodotto scalare, 
\begin{equation}
\langle\, \psi_{s}\, ,\, {\bm \psi}\, \rangle\equiv
 \int \psi_{s}^\dagger(\vec{x}) \, {\bm \psi}(x)\, d^3x= 
{\bm a}_s(t)
\end{equation}
un'espressione 
nella quale gli indici spinoriali sono contratti secondo  $\displaystyle\xi^\dagger \lambda=\sum_{a=1}^4 \xi^*_a \lambda_a$. 

Nel caso speciale delle particelle libere avremo
 \begin{equation}\label{borno}
 \psi_{s}(\vec{x})=\frac{u_s}{\sqrt{2\, \varepsilon_s \, V}} \,  e^{i(\vec{p}\, \vec{x})/\hbar} 
 \mbox{ e } {\bm a}_s(t)={\bm a}_s(0)\, e^{\mp  i(\varepsilon_s t)/\hbar}
 \end{equation}
dove i segni $-$ e $+$ indicano  gli stati con energie positive e negative, rispettivamente,  e 
 dove $V$ \`e il volume formale della porzione di spazio tridimensionale 
sul quale ci basiamo per effettuare la quantizzazione. 
Anche 
se la funzione~\ref{borno} \`e normalizzata {\em \`a la Born}, 
abbiamo preferito utilizzare i  comuni 
4-spinori $u_s$ che soddisfano $u_s^\dagger u_s=2\varepsilon_s$
per facilitare i riferimenti coi testi moderni.

Per descrivere degli elettroni, si assume che gli operatori ${\bm a}_s$ abbiano natura fermionica, ovvero, anti-commutante, e lo stesso vale per i campi 
${\bm \psi}$. Proprio come nel caso della presentazione in sez.~\ref{s-ond},  
si tratta di una posizione semplicemente giustapposta al resto del costrutto teorico. 
Ribadiamo il punto con un commento a proposito:  lo spin \`e automaticamente incluso una volta che usiamo 
 l'equazione di Dirac come equazione d'onda, mentre 
 il carattere statistico dei fermioni  viene invocato come  ipotesi addizionale per non contraddire i fatti noti.

\section{Majorana e la quantizzazione canonica dei fermioni}\label{s-maj}
Nel 1937 Majorana propone una nuova procedura di quantizzazione \cite{m37}. Egli inizia osservando che l'equazione di Dirac per una particella 
libera consentirebbe di trattare anche il caso di una funzione d'onda reale, tale che $\psi^*=\psi$.
Tuttavia, la  relativa densit\`a di energia, simile a quella che si costruisce per il campo elettromagnetico,\footnote{Majorana giustifica questa densit\`a a partire della densit\`a invariante di lagrangiana, che corrisponde all'equazione di Dirac. 
In queste note introduciamo il coefficiente $\nicefrac{1}{2}$ solo per uniformarci alle notazioni moderne; ovviamente volendo 
lo possiamo assorbire nella funzione $\psi$, come fa Majorana.  Inoltre, in questa sezione esibiamo per massima chiarezza gli indici spinoriali delle funzioni d'onda e dei campi, indicati con $a,b,c...$, 
nella speranza che questo non induca confusione con gli indici $s,s'$, che  usiamo altrove per indicare i numeri quantici che identificano gli stati.} 
\begin{equation}
 \frac{1}{2} \psi^t \Delta \psi=
\frac{1}{2} \psi_a \Delta_{ab} \psi_b \qquad
\mbox{ dove }a,b=1,2,3,4
\end{equation}
fornisce una energia pari a zero.\footnote{\`E curioso che Pauli ripeta l'osservazione di Majorana  quattro anni dopo~\cite{p41} senza concedere credito al collega.} 
La verifica \`e diretta: 
 il termine  in $\Delta$ proporzionale alla massa contiene la matrice asimmetrica $\beta$; 
 l'altro termine si riduce ad una derivata totale, siccome,  ricordando che le matrici $\alpha$ sono simmetriche, 
 $ 2 \psi_a \vec{\alpha}_{ab} (\vec{\nabla} \psi_b) = \vec{\nabla}(\psi \vec{\alpha}  \psi  ) $.  Pertanto, assumendo che le funzioni d'onda degli stati abbiano 
 valori  trascurabili al contorno,
 arriviamo alla conclusione anticipata.  Ma perch\'e mai non dovrebbe essere possibile 
 descrivere un'onda di tipo reale, quando l'equazione differenziale che la descrive \`e reale? 
 Sembra utile pensarci pi\`u a fondo.
  
 \subsection{L'ipotesi dei campi hermitiani e una deduzione di Majorana}\label{buste}
Majorana  procede dichiarando 
 che non stiamo trattando una funzione d'onda, bens\'i un operatore: ovvero, suggerisce di definire un opportuno 
 ``campo quantizzato'', nome che richiama il contesto di discussione: quello della fisica.
 Per evitare di confondere l'operatore con una funzione numerica, proprio come abbiamo fatto in precedenza,
 useremo un carattere in grassetto: 
 \begin{equation}\psi \to {\bm \psi}\end{equation}
 Con questa considerazione in mente
 dobbiamo far attenzione a non assumere {\em a priori} che tale oggetto matematico 
 sia commutante - e come vediamo qui di seguito, Majorana adduce ottime ragioni per assumere che non lo sia. 
 
 Supporremo che l'hamiltoniana del campo hermitiano ${\bm \psi}$,  
 che andiamo a specificare, sia il corrispettivo di quella appena esibita 
 \begin{equation}
 {\bm H}=  \frac{1}{2}  \int {\bm \psi}_a(x) \  (\Delta {\bm \psi}(x) )_a\ d^3 x = 
 -\frac{1}{2}  \int ({ \Delta \bm\psi(x)})_a \   {\bm \psi}(x) _a\ d^3 x \label{mplz}
 \end{equation}
Le manipolazioni  mostrate appena sopra permettono di spostare l'operatore differenziale $\Delta$ sul campo a sinistra,
 cambiando il segno dell'integrale e scartando il termine di derivata totale.\footnote{ Seguendo Majorana, possiamo riassumere le manipolazioni in eq.~\ref{mplz} dicendo che  
 l'operatore $\Delta$, che agisce sullo spazio 3 dimensionale e sugli indici spinoriali, \`e antisimmetrico;
 si ricordino le propriet\`a  di eq.~\ref{pm}.}
 
Nel passaggio cruciale,  Majorana richiede che l'equazione differenziale coincida con l'equazione di moto di Heisenberg
\begin{equation}
i\hbar \frac{d}{dt} {\bm \psi}_b(y)=[ {\bm \psi}_b(y)\,  , \, {\bm H}]
\end{equation}
dove a destra abbiamo un commutatore: $[\bm{X},\bm{Y}]=\bm{X}\, \bm{Y}- \bm{Y}\, \bm{X}$.
Siccome l'hamiltoniana \`e indipendente dal tempo, 
faremo in modo che tutti i campi siano calcolati 
allo stesso tempo $t$.

Nel membro di destra, 
sommiamo e sottraiamo  il termine 
\begin{equation}
\frac{1}{2}  \int {\bm \psi}_a(x)\  {\bm \psi}_b(y)\  (\Delta{\bm \psi}(x))_a \ d^3 x = 
-\frac{1}{2}  \int (\Delta {\bm \psi}(x) )_{a}\ {\bm \psi}_b(y)\  {\bm \psi}_a(x)\ d^3 x 
\end{equation}
che, come appena mostrato, si pu\`o riscrivere spostando  $\Delta$ sul campo a sinistra (dopo aver scartato una derivata totale) siccome tale operatore 
differenziale agisce  solo sulle coordinate di integrazione $x$ e non sulle $y$.
In questo modo,  si perviene alla condizione di coerenza
 \begin{equation}
i\hbar \frac{d}{dt} {\bm \psi}_b(y)=   \frac{1}{2} \int \Big(  C_{ab}(x,y) \ (\Delta {\bm \psi}(x)  )_a
+  (\Delta {\bm \psi}(x) )_a  \  C_{ab}(x,y)  \Big) d^3 x \label{20}
\end{equation}
con 
\begin{equation}
C_{ab}(x,y)= \{ {\bm \psi}_b(y) , {\bm \psi}_a(x)  \} 
\end{equation}
dove a destra  
abbiamo un anti-commutatore: $\{\bm{X},\bm{Y}\}=\bm{X}\, \bm{Y}+ \bm{Y}\, \bm{X}$.
La condizione  di coerenza pu\`o essere immediatamente implementata ponendo 
\begin{equation}
\{ {\bm \psi}_b(y) , {\bm \psi}_a(x)  \} = \delta^3(\vec{x}-\vec{y})\, \delta_{ba}
\end{equation}
che mostra come i campi introdotti da Majorana e 
regolati dall'equazione di Dirac 
sono anti-commutanti,  una 
propriet\`a che (nel formalismo di Jordan, Klein, Wigner, Heisenberg e Fock) caratterizza i campi fermionici.
Curiosamente, questa essenziale osservazione \`e poco ricordata.

 \subsection{Espansione in oscillatori}\label{espa}
 Per completezza didattica e per confrontarci bene con la sezione~\ref{vedq}, espandiamo esplicitamente il campo hermitiano come segue  
\begin{equation}
{\bm \psi}=\sum_s  \left( {\bm a}_s\,  \psi_{s} + {\bm a}_s^\dagger\, \psi_{s}^* \right) \label{qc}
\end{equation}
dove gli stati $s$ hanno  energia positiva, vengono identificati dal momento 
e dall'elicit\`a e sono tra loro ortogonali (si confronti con l'eq.~\ref{sq}).
Valgono ancora relazioni tipo le precedenti
\begin{equation}
{\bm a}_s = \langle\,  \psi_{s} \,  ,\,  {\bm \psi} \, \rangle 
\mbox{ e } 
{\bm a}_s^\dagger = \langle\,  \psi_{s}^* \,  ,\,  {\bm \psi} \, \rangle 
\end{equation}
dove ricordiamo che le funzioni d'onda coniugate sono soluzioni dell'equazione libera, ortogonali alle altre,\footnote{La dimostrazione \`e molto semplice. Infatti, prendendo le onde libere, 
autofunzioni dell'equazione differenziale, basta osservare che $\Delta \psi_s=\varepsilon_s \psi_s$, dove $\varepsilon_s\neq 0$ (eq.~\ref{essoeps}).
Dalle considerazioni all'inizio di questa sezione, segue che $\psi_a\psi_a=\psi^t \psi=0$, che \`e la condizione di ortogonalit\`a tra una funzione d'onda e la sua coniugata. } e dove 
gli indici spinoriali sono tra loro contratti. 
Dalla condizione sull'anti-commutatore, troviamo  
\begin{equation}
\begin{array}{c}
 \{ {\bm a}_s  \, ,\,  {\bm a}^\dagger_{s'}  \}  = 
 \{\,  \langle \psi_{s} \ , {\bm \psi} \rangle \ ,\  \langle \psi_{s'}^*, {   \bm \psi}  \rangle\,   \} = \\[1ex]
 \int   (\psi_{s}^*(\vec{x}))_a  \ (\psi_{s'}(\vec{y}))_b  \times 
 \{  \, {\bm \psi}_a (x)     ,   {\bm \psi}_b(y) \, \} \    d^3x\, d^3y=  \\[1ex]
 \int   (\psi_{s}^*(\vec{x}))_a  \  (\psi_{s'}(\vec{y}))_b   \times      \delta^3(\vec{x}-\vec{y}) \ \delta_{ab}  \ 
d^3x\, d^3y= \\[1ex]
\int \psi_{s}^\dagger (\vec{x})\, \psi_{s'}(\vec{x})\, d^3x= \langle \psi_{s} \ ,  \psi_{s'} \rangle =\delta_{s's'}
\end{array}
\end{equation} 
ad ogni tempo $t$.  L'ortogonalit\`a della funzione d'onda e della sua coniugata implica che gli altri anticommutatori sono nulli. 
 
 \subsection{Quantizzazione dell'elettrone}\label{este}
 Per prima cosa, osserviamo che grazie a Pauli e Majorana si perviene alla conclusione (anticipata dalle intime convinzioni dei due scienziati) che il mare di Dirac 
 sia un costrutto non solo poco attraente, ma anche non necessario; o per dirla in modo succinto, che esso non esista.\footnote{Questo non significa che esso non abbia svolto un 
 ruolo  propulsivo nel costruire  la prima teoria delle interazioni deboli, incluse la previsione della cattura elettronica 
 e del decadimento $\beta^+$ \cite{wi34} come sopra ricordato; vedi~\cite{quad} per maggiore discussione.
Il mare di Dirac potrebbe ancora essere d'ausilio  per {\em iniziare} a parlare di anti-materia.}

\`E notevole che l'argomento di Majorana porti a ritenere che  si debba applicare 
la statistica di Fermi-Dirac alle particelle di spin  $\nicefrac{1}{2}$; un aspetto non secondario della 
connessione tra spin e statistica,  non  sempre agevole da apprezzare nei percorsi didattici. 
Un punto altrettanto importante della sua teoria riguarda la quantizzazione di un campo non hermitiano, per il quale gli stati di particella e di anti-particella sono distinti dalla carica elettrica, come nel caso notevole dell'elettrone;
per far questo basta considerare \cite{m37} la seguente ovvia costruzione
\begin{equation}
{\bm \psi}=\frac{{\bm \lambda}+i\, {\bm \chi}}{\sqrt{2}} \label{qce}
\end{equation}
dove $\bm\lambda$ e $\bm\chi$ sono i campi quantizzati hermitiani gi\`a definiti, che descrivono particelle con la stessa massa,  
dalla cui anticommutazione segue 
\begin{equation}
\{ {\bm \psi}_b^\dagger(y) , {\bm \psi}_a(x)  \} = \delta^3(\vec{x}-\vec{y})\ \delta_{ba}
\end{equation}
Il campo quantizzato complessivo pu\`o essere scritto come
\begin{equation}
 {\bm \psi}=\sum_s  \left(   
  {\bm c}_s   \,  \psi_{s} + 
 \bar{\bm c}_s^\dagger  \, \psi_{s}^* \right) 
 \end{equation}
 con ${\bm c}_s\neq \bar{\bm c}_s$, dove
 \begin{equation}
{\bm c}_s=  \frac{{\bm a}_s + i  {\bm b}_s }{\sqrt{2}} 
\quad \mbox{ e } \quad 
\bar{\bm c}_s=  \frac{ {\bm a}_s -  i  {\bm b}_s }{\sqrt{2}} 
 \end{equation}
L'hamiltoniana libera di ${\bm \psi}$  coincide con due hamiltoniane libere indipendenti 
\begin{equation}
  \int {\bm \psi}^\dagger(x)  \  \Delta {\bm \psi}(x)  \ d^3 x 
 = \frac{1}{2}\int {\bm \lambda}(x)  \  \Delta {\bm \lambda}(x)  \ d^3 x +
  \frac{1}{2}\int {\bm \chi}(x)  \  \Delta {\bm \chi}(x)  \ d^3 x 
\end{equation}
L'inclusione delle interazioni elettromagnetiche, ottenuta rimpiazzando $\Delta \to \Delta_q$, conduce a termini misti 
tra questi due campi, ovvero, al loro `mescolamento'. Si arriva alla cosiddetta 
hamiltoniana di Dirac, ma scritta utilizzando il campo quantizzato ${\bm \psi}$.\footnote{\`E agevole
verificare l'invarianza di gauge dei vari termini corrispondente a
$\delta{\bm \psi}=i \epsilon {\bm \psi}\Leftrightarrow(\delta{\bm \lambda}=- \epsilon {\bm \chi} \, ,\, \delta{\bm \chi}=+ \epsilon {\bm \lambda} )$.}

\section{Discussione del caso dei fermioni neutri}\label{s-sce}
Gli sforzi di Dirac, rivolti a pervenire ad  una teoria dell'elettrone basandosi sulla sua equazione d'onda, 
lo conducono  nel 1931 a proporre la procedura descritta 
in sez.~\ref{s-dir}.
Tuttavia,  due nuove particelle fermioniche, 
prive di carica elettrica,  si impongono  alla considerazione degli scienziati negli stessi anni.   
In effetti, nel 1932  Chadwick scopre il neutrone, che Iwanenko e Heisenberg mostreranno 
essere l'ingrediente mancante per il modello del nucleo dell'atomo (1933). Inoltre 
il concetto di neutrino, proposto da Pauli nel 1930, viene modificato e portato a completa 
maturazione prima da Fermi (1933) e poi da Majorana (1937). Come si confronta la descrizione  dei neutrini e dei neutroni 
con quella degli elettroni?

1)~Nella teoria di Fermi del decadimento $\beta$ (1933) si ipotizza  
l'emissione associata di elettroni e neutrini relativistici, descritti seguendo le idee proposte da 
Dirac \label{vedq} (sez.~\ref{vedq}).
Dunque, Fermi si trova a postulare che esista un mare di Dirac per gli elettroni, ed uno per i neutrini. 
Cos\`i facendo, fermioni e anti-fermioni risultano ben distinti tra di loro; vale per l'elettrone e vale per il neutrino \cite{quad}. 
Si tratta insomma proprio del caso che oggi va sotto il nome di  ``neutrino di Dirac''.

2)~La teoria di Majorana  offre  possibilit\`a nuove. Mentre   una particella dotata di carica elettrica {\em deve} essere 
ben distinta dalla sua anti-particella (e quindi richiede un campo non-hermitiano come quello descritto in sez.~\ref{este}) nel caso di fermioni neutri esiste una possibilit\`a pi\`u semplice: quella 
che siano descritti da un campo hermitiano, come quello discusso in sez.~\ref{buste}. Questo significa considerare particelle 
a spin $\nicefrac{1}{2}$, fermioniche,   identiche alle proprie antiparticelle. 
Majorana propose che questa semplificazione  formale si potesse applicare a neutrini ed a neutroni \cite{m37}. Ma pochi mesi dopo, Racah osserv\`o 
che le particelle di  Majorana devono esser prive non solo di carica elettrica, ma anche di momento magnetico \cite{g37},\footnote{Gli indizi che il neutrone \`e dotato di un (grande)  momento magnetico erano emersi sin dalla met\`a degli anni 30.} ragion per cui   
non crediamo che la proposta di Majorana si possa applicare ai neutroni cos\`i come li conosciamo.\footnote{Il neutrone \`e distinto dall'anti-neutrone dalla carica barionica, ipotizzata per dare conto della stabilit\`a della materia;   ma d'altro canto, almeno a livello  speculativo, essa potrebbe essere debolmente violata, e in questo caso gli autostati di massa nel sistema a riposo non sarebbero un neutrone ed un anti-neutrone con massa uguale, ma due particelle di Majorana con masse lievemente diverse. In questo caso, 
neutroni ed anti-neutroni  potrebbero ``oscillare'' tra di loro, proprio come capita
ai mesoni $K^0-\bar{K}^0$, ai neutrini, ecc.~\cite{p57}.} Invece, l'idea che i neutrini siano particelle di Majorana \`e considerata 
promettente ed \`e oggi vivacemente investigata: vedi p.e.~\cite{what21,ago}.

3) Consideriamo infine il trattamento di particelle ed antiparticelle,  
che  tra quelli qui considerati \`e l'ultimo ad essere proposto, e che si basa su una re-interpretazione dell'equazione di Dirac. 
Pur non essendo del tutto soddisfacente,  esso ha il vantaggio di essere  molto semplice (sez.~\ref{s-ond}). Inoltre, 
esso \`e {\em compatibile} con la proposta di Majorana: usando il formalismo della sez.~\ref{s-ond} \`e sufficiente porre
$\psi_+=\psi_-$. Naturalmente, perch\'e le due equazioni d'onda coincidano la loro 
carica elettrica deve essere nulla, vedi eq.~\ref{pazz}, cos\`i come il loro momento magnetico anomalo.  
Per una discussione un po' pi\`u articolata, sviluppata nello stesso spirito della presente, si veda nuovamente~\cite{what21}.

\section{Un riassunto del percorso storico}\label{s-sto}
Dopo aver esposto e discusso vari formalismi per descrivere i campi fermionici quantizzati, in un modo che trovo pi\`u ordinato e  
meno impegnativo di quanto si ritenga di solito necessario, 
provo a riassumere il percorso storico che si snoda dalla fine degli anni 20 fino all'inizio degli anni 40.
Ecco i principali passi avanti verso la moderna comprensione dell'anti-materia:
{\small
\begin{itemize}
\item Nel 1928 Dirac \cite{d28} scopre l'eponimica equazione, dimostrando: 
1)~che essa implica l'esistenza dello spin dell'elettrone e 2)~che ne prevede  il corretto momento magnetico.
\item Dotandola di addizionali ipotesi per consentire una interpretazione  diretta e coerente con le osservazioni, 
egli propone che esistano gli {\em anti-elettroni} \cite{d31}, ovvero dei costrutti teorici che discendono dall'ipotesi dell'esistenza del mare di Dirac (1931).
\item Anderson scopre il {\em positrone} \cite{a32} e lo presenta come un elettrone positivo (1932).
\item Nel 1933 i contributi di Dirac sono generalmente riconosciuti. La sua lezione per il premio Nobel in fisica si intitola 
\begin{quote}
{\em Theory of Electrons and Positrons,}
\end{quote}
a rivendicazione del fatto che il concetto di anti-elettrone
della sua teoria sia sovrapponibile al positrone appena osservato.
\item Prima Klein \& Nishina \cite{kn29}, poi Fermi \cite{f33} e dopo la scoperta di Anderson molti altri teorici~\cite{quad}
usano la teoria di Dirac per  derivare previsioni. 
La prima forma di campo quantizzato
per gli elettroni,    detta  campo di ``seconda quantizzazione'' e simile al formalismo 
che si  usa ancora oggi in contesti non-relativistici, 
si basa sull'ipotesi dell'esistenza del mare di Dirac e viene utilizzata almeno fino al 1937; a volte anche in seguito.
\item Pauli e Weisskopf  \cite{pw34} mostrano come produrre una interpretazione  coerente 
per una particella ipotetica e senza spin, che non richiede il mare di Dirac (1934). Entrambi procedono nelle loro indagini in varie direzioni valide \cite{v35,p36-1,p36-2}, ma non quantizzano le particelle con spin $\nicefrac{1}{2}$.
\item Finalmente, Majorana  \cite{m37} mostra  come  si possano trattare elettroni e positroni in uno contesto 
teorico che non abbisogna del mare di Dirac (1937); si tratta  proprio 
del formalismo della quantizzazione canonica per le particelle con spin $\nicefrac{1}{2}$, presentato con la scelta di matrici di Dirac adottata  nella presente nota. Egli dimostra che i campi anti-commutano, dunque 
tali particelle devono obbedire alla statistica di Fermi Dirac e rispettare il principio di esclusione. 
La stessa conclusione vale per i neutrini, per i quali Majorana prospetta nuove possibilit\`a teoriche.
\item Pauli 1941 \cite{p41} stila un'influente rassegna che riassume, in parte riformulando, la discussione sui campi quantizzati.
\item Stueckelberg  1941 \cite{s41,s42} evidenzia il punto di vista della meccanica ondulatoria, con un lavoro visionario che preparer\`a il successivo e ben noto contributo di Feynman. 
\end{itemize}}
Purtroppo, gli ultimi due autori non danno pieno riconoscimento al precedente contributo Majorana,  del quale oggi \`e evidente  
il carattere  fondativo
(vedi anche \cite{quad,majo} e 
 l'appendice~\ref{appb}).
Questo render\`a pi\`u lenta e meno completa  la successiva elaborazione del concetto di campo quantizzato, con effetti particolarmente deprecabili
sulla  didattica che temo perdurino ai giorni nostri; anzi, \`e proprio tale considerazione che mi ha motivato a stilare la presente nota.

L'opinione di Fermi sul risultato di Majorana \`e molto pi\`u lusinghiera di quella che ci trasmettono i lavori di Pauli
\cite{r1,r2} anche se non si traduce in una presa di posizione pubblica (vedi di nuovo  \cite{majo,quad}). 
Il riassunto del lavoro del 1937, che riportiamo qui di seguito,  
si riferisce ai precedenti fatti di storia culturale e non lascia molti 
dubbi sul suo contenuto:
\begin{quote}
\small
Si dimostra la possibilit\`a di  pervenire a una piena simmetrizzazione
formale della teoria quantistica dell'elettrone e del positrone facendo
uso di un nuovo processo di quantizzazione. Il significato delle
equazioni di Dirac ne risulta alquanto modificato e non vi \`e pi\`u luogo
a parlare di stati di energia negativa; n\`e a presumere per ogni altro
tipo di particelle, particolarmente neutre, l'esistenza di \guillemotleft antiparticelle\guillemotright\ 
corrispondenti ai \guillemotleft vuoti\guillemotright\ di energia negativa.
 \end{quote}
Majorana  illustra la possibilit\`a
di  eliminare dalla teoria il  mare di Dirac, 
evitando ogni riferimento alle energie negative.
Il suo argomento, esposto con minimi adattamenti in 
sez.~\ref{s-maj}, introdotti per fini didattici,   ci porta ad ammettere che le particelle  
descritte dall'equazione di Dirac si comportino come fermioni.
L'uso di una particolare scelta di matrici $\gamma$ \`e solo un modo di semplificare la discussione,
e l'applicazione alle particelle neutre \`e una conseguenza naturale della procedura di quantizzazione.
Si noti l'enfasi data alle particelle neutre  nel riassunto, che si riferisce alle considerazioni 
esposte in sez.~\ref{s-sce} e che di solito \`e l'unico risultato che viene oggi ricordato
del lavoro di Majorana.

\subsection{Annotazioni}\label{azioni}

Grazie al percorso intellettuale qui riassunto,  che si sviluppa in una dozzina di anni, 
viene messo a punto un affidabile e pratico modello  teorico  che descrive la coppia elettrone-positrone. 
Da qui si arriva presto a ragionare  di materia e di anti-materia per poi  approdare 
al concetto ancora pi\`u generale di particella ed antiparticella, con un formalismo che descrive  luce e materia in modo simile. 
Credo tuttavia che ci sia molto valore didattico nell'iniziare facendo presenti le distinzioni 
tra materia e radiazione, tra elettroni e neutrini, tra nucleoni e nuclei: 
passando dall'insegnamento della fisica atomica  a quello della fisica nucleare,  o a quello delle particelle, \`e bene 
prendersi un attimo per sottolineare senza dare per scontati   gli aspetti  che caratterizzano i vari ambiti.

Il  dibattito sulla  struttura fondamentale della 
 materia eredita dal passato  certe aspettative:  p.e., 
 si parte pensando che la materia e la luce siano sostanze radicalmente diverse;  
  l'atomismo induce  inizialmente a 
credere che le parti minime della materia siano permanenti. Alcune di queste aspettative 
   si evolvono nel corso del tempo, altre ancora vengono contraddette; ma, per l'appunto, occorre del tempo per 
   pervenire a tali sviluppi.
 Noto poi che la discussione  in merito al tema della materia 
 non sembra del tutto conclusa: penso alla questione dell'origine del numero barionico ed  
 al neutrino di Majorana, particella ipotetica a cavallo tra i mondi della materia e dell'anti-materia.

Due annotazioni importanti in vista di sviluppi successivi della discussione:
\begin{itemize}
\item  Pauli, Weisskopf  e Majorana sviluppano le loro argomentazioni evidenziando il concetto di azione,
utilizzato occasionalmente in precedenza nelle discussioni della meccanica ondulatoria, 
che diventer\`a assolutamente centrale 
per la procedura di quantizzazione di Feynman~\cite{f49}. 
L'unica 
ragione per cui   ho  dato risalto all'hamiltoniana e non  all'azione  \`e il carattere  introduttivo di questa proposta: 
nei corsi introduttivi di meccanica quantistica non relativistica si usano quasi esclusivamente
delle hamiltoniane.
Ma le densit\`a 
di lagrangiana sono importanti  ed evidenziano il parallelismo tra i modi in cui 
trattiamo i campi di radiazione e quelli di materia;   nei percorsi didattici avanzati \`e assai opportuno (o persino necessario) parlarne.
\item Di nuovo   Pauli, Weisskopf  e Majorana danno grande evidenza 
al concetto di campo quantizzato (scalare e spinoriale), inaugurando di fatto la moderna 
teoria quantistica dei campi. Ricordiamo per\`o che nel metodo di quantizzazione di 
Stueckelberg-Feynman
 (dove si sostituisce il concetto di energia negativa con quello di propagazione all'indietro nel tempo) 
    si evita di ricorrere a questo costrutto teorico. 
Per tenere conto del carattere fermionico delle particelle con spin $\nicefrac{1}{2}$
si usano i cosiddetti numeri anti-commutanti o di Grassmann.
 \end{itemize}

\section{Discussione}\label{s-dis}

\begin{table}[t]
\begin{center}
\small
{\begin{tabular}{| l|c| |  c | r | }
\hline
\sc Maxwell  & 
Teoria dell'elettromagnetismo  &  
Teoria dei buchi  & 
 \sc Dirac  \\[-0.2ex]
  1873 & 
con interpretazione meccanica &  
basata sul mare di Dirac & 
  1931 \\ \hline
\sc Hertz & 
Produzione di&  
Scoperta& 
 \sc Anderson   \\[-0.2ex] 
1880 & 
onde che si propagano &  
del positrone  & 
 1932  \\
\hline
\sc Einstein   & 
Accantonamento  &  
Quantizzazione canonica & 
\sc Majorana   \\[-0.2ex]
 1905
   & 
dell'ipotesi dell'etere  &  
dei fermioni & 
   1937 \\ \hline
\end{tabular}}
\end{center}
\caption{\em\small Un parallelo tra i destini di due concetti: 1)~quello di etere, originariamente introdotto per ragionare sulle onde luminose, 
e 2)~quello di mare di Dirac, introdotto per ragionare sulle onde degli elettroni nel contesto della teoria relativistica, senza contraddire l'esistenza degli atomi.
}\label{tab2}
\end{table}

 \subsection{Commenti e sommario degli spunti proposti}

In dipendenza dalle esigenze concrete della didattica (incluso il tempo a disposizione) 
potrebbe essere  interessante e persino istruttivo considerare  l'idea di esporre 
la strada originariamente seguita da Dirac non solo come esempio di teoria oggi abbandonata, ma quale utile `artificio', che \`e il termine 
che Fermi, nella prima teoria dell'emissione dei raggi $\beta$, scelse per qualificarla \cite{f33}. Va aggiunto che, se si evitano questi
passi, molti lavori dell'epoca, inclusi quelli di Fermi~\cite{f33} e di Majorana~\cite{m37} diventano  quasi illeggibili per un lettore moderno; 
si veda \cite{quad} per maggiore discussione.

Adottando in modo coerente
 il formalismo della funzione d'onda e accompagnandolo con l'interpretazione esposta in sez.~\ref{s-ond}
\`e agevole  anticipare la nozione  di anti-particella nei corsi introduttivi (volendo farlo); 
per questo l'ho presentata per prima. 
Si deve solo  fare attenzione  a evidenziare l'importanza di una re-interpretazione del 
senso delle funzioni d'onda coniugate, ed essere coscienti che il carattere statistico delle particelle 
{\em non emerge} da queste considerazioni. 

L'approdo finale della discussione, dovuto a Majorana, \`e molto economico ed 
indica una completa coerenza tra il quadro teorico ed i fatti osservativi.  
Le particelle di materia, quantizzate {\em \`a la Majorana,}
non possono essere trattate come onde classiche. La procedura indica in modo trasparente che le propriet\`a statistiche degli elettroni 
sono ben diverse da quelle dei fotoni. 
L'originaria nozione di anti-materia (di Dirac) acquisisce un nuovo significato, che possiamo
 adottare con convinzione e coerenza per descrivere le nuove procedure di quantizzazione, 
ridefinendo i precedenti schemi concettuali.  
Propongo due osservazioni per chiarire meglio questa considerazione:\\
$\bullet$ il processo che oggi viene denominato ``creazione di coppie'', evidenziando l'apparizione di particelle precedentemente assenti, 
non comporta {\em alcuna} creazione di particelle nel contesto della teoria dei buchi, ma solo
una dislocazione di un elettrone del mare di Dirac tra gli stati di energia positiva. Si veda~\cite{libro} per maggiori discussioni. \\
$\bullet$ La definitiva eliminazione del mare di Dirac  dalla teoria somiglia un po' all'accantonamento del concetto di etere.\footnote{Per capirne la 
perdurante influenza basti  un esempio: 
 ancora nel 1950, il famoso lavoro di Foldy e Wouthuysen \cite{fw}
parla di energie negative, ed il celeberrimo libro di Dirac \cite{paulo}, tuttora molto usato 
nelle universit\`a,  presenta nel \S~65  il costrutto concettuale di ``buco'' e  nel \S~73 
quello di ``mare di Dirac'', pur senza usare questa ultima denominazione.} 
La tabella~\ref{tab2}, adattata da \cite{majo}, illustra le pi\`u significative analogie 
ed elenca i pi\`u importanti passi avanti. \\
Sottolineo che alcuni dei pi\`u  importanti progressi concettuali, e specialmente quelli teorici, 
 non possono che  sfuggire a chi non si soffermi almeno un istante a ponderarne il valore.

 \subsection{Osservazioni critiche e valutazioni}\label{pimp}
 
Il difetto dei percorsi didattici che evidenziano la storia delle idee (come il qui presente)   \`e quello di dover 
ripetere dei passi faticosi. Questo pu\`o suggerire di evitare di farlo, del tutto o almeno in parte. 
D'altro canto, offrire immediatamente la conclusione di una discussione durata pi\`u di un decennio, senza accennare ai punti dibattuti o senza evidenziare 
le scelte fatte, pu\`o risultare altrettanto penoso per uno studente dotato di senso critico. 

P.e., l'espressione del campo quantizzato fermionico di solito presentata \`e qualcosa del genere
\begin{equation}
{\bm\psi}(x)=\sum_{\lambda=\pm\nicefrac{1}{2}}
\int \frac{d^3p}{(2\pi)^3 2 \varepsilon_{\vec{p}}} \left( \   \tilde{\bm  a}_{\vec{p}\lambda}\, \tilde{u}_{\vec{p}\lambda}\, e^{i\, px}   + 
\tilde{\bm  b}_{\vec{p}\lambda}^\dagger\, \tilde{v}_{\vec{p}\lambda}\, e^{-i\, px}   
\ \right) \label{cov}
\end{equation}
Perch\'e uno studente la capisca   deve padroneggiare, oltre alle soluzioni libere dell'equazione di Dirac e agli spazi di Fock: 
il sistema di unit\`a naturali $\hbar=c=1$;
la matrice di coniugazione di carica, implicita nella espressione degli spinori $v$ ed esaminata a fondo dopo Majorana \cite{c}; 
le notazioni Lorentz covarianti per i 4 vettori, per la misura d'integrazione, per 
la normalizzazione degli oscillatori e per  quella degli spinori.\footnote{Ho incluso la tilde in eq.~\ref{cov} per
evidenziare questi due ultimi aspetti, evitando la confusione con gli  spinori in eq.~\ref{borno} e gli oscillatori in eqq.~\ref{sq}, \ref{qc} o \ref{qce}.}
In un percorso didattico, questi punti devono essere stati discussi accuratamente in precedenza e  connessi con il senso dell'espressione.
Ad onor del vero, {\em nessuno} di tali aspetti della presentazione, oggi ritenuti importanti se non essenziali, compare 
nei lavori originari, nei quali si percepisce invece lo sforzo di  far risaltare il senso fisico e di 
far corrispondere   il vecchio formalismo  
col nuovo, proprio come nell'esposizione delle sezz.~\ref{vedq}, \ref{squ}, \ref{espa} e \ref{este}.
Le prime  sofisticazioni di tale tipo compaiono  in una equazione non numerata della rassegna di Pauli \cite{p41} a pagina 224.

La presenza di un operatore coniugato in eq.~\ref{cov}
 sembra essere un dettaglio tra i tanti quando \`e il nucleo
della quantizzazione canonica. 
Le espressioni del campo di seconda quantizzazione  di Fermi (eq.~\ref{sq}),  
e del campo di Majorana  (eq.~\ref{qc} e \ref{qce}),   risultano  a confronto   cos\`i succinte
da mettere a disagio chi sia gi\`a abituato all'eq.~\ref{cov}. Sono mica sbagliate?  
Sono un'altra cosa? vien da chiedersi.

Nelle presentazioni usuali
il campo ${\bm \psi}$ 
viene definito 
a partire dagli operatori ${\bm a}$ e  ${\bm b}$, che \`e il contrario della procedura di Majorana 
descritta sez.~\ref{espa};  questo toglie immediatezza al concetto di campo quantizzato fermionico e mi sembra faccia sentire come artificiale la sua costruzione. 
Inoltre, nonostante l'eq.~\ref{cov} (o analoghe)  descriva un operatore nello schema di Heisenberg (o di interazione) nel lato di destra compare
il fattore $ e^{-i\, px}$, che suggerisce che siano le funzioni d'onda che cambiano nel tempo; un ulteriore potenziale  ragione di confusione per un novizio. 
Infine, si noter\`a che l'eq.~\ref{cov} 
d\`a  una grande enfasi al caso speciale dei 
campi liberi, circostanza che  non mi sembra contribuisca alla chiarezza concettuale.
A ben considerare,  lo stesso Fermi si preoccupa sin dal 1933 
di includere gli effetti del campo elettromagnetico del nucleo per le funzioni d'onda  degli elettroni emessi nel decadimento $\beta$ \cite{f33}.

\paragraph*{Ringraziamenti}
{\footnotesize Lavoro 
col parziale supporto della borsa di ricerca 2022E2J4RK {\em PANTHEON: Perspectives in Astroparticle and Neutrino THEory with Old and New Messengers,}
 nell'ambito del programma PRIN 2022, finanziato dal  ``Ministero dell'Universit\`a e della Ricerca'' (MUR) e basato  sui
 seminari del 16 gennaio 2025 presso l'universit\`a di Amburgo, 
 del 25 febbraio 2025 presso l'universit\`a di Roma Tor Vergata, del 14 marzo 2025 presso la
 Scuola superiore ``Carlo Urbani''  dell'universit\`a di Camerino.
Ringrazio  Salvatore Esposito, Roberto Lalli, Orlando Luongo, Guenter Sigl, Andreas Ringwald, e Nazario Tantalo 
per preziose discussioni e un anonimo revisore del Giornale di Fisica per una {\em utilissima} serie di osservazioni. 

}

\appendix

 \begin{table}[t]
 \small
\centerline{\begin{tabular}{|lc|ccc|}
\hline
Autore & data & riferimenti &   riferimenti alla quant- &  riferimenti \\
e referenza &   &  precedenti  &   tizzazione del 1937 \cite{m37}  &  successivi \\
\hline
Heitler \cite{h36} & 1936 & \cite{jw}, \cite{p36-2} & - & - \\
Wentzel \cite{w43} & 1943 & \cite{jw},\cite{pw34} & no & no \\
Schweber \cite{s61} & 1961  & \cite{jw},\cite{pw34}  & no & \cite{p40}\\ 
Bjorken, Drell \cite{bd65} & 1965 & \cite{jw} & no & \cite{p40}\\ 
\hline\hline
Brown, Hoddeson  \cite{b83} & 1983  &\cite{jw},\cite{pw34} & no & \cite{p41} \\
Pais \cite{p86} & 1986  & \cite{jw},\cite{pw34} &  (neutrino, $\gamma^*_\mu=-\gamma_\mu$) & no \\
Schweber  \cite{s94} &  1994  & \cite{jw},\cite{pw34}   &  (lezione di Schwinger) & \cite{p40},\cite{p41} \\
Jacob {\em et al}  \cite{mj} & 1998 &  \cite{pw34}   &  si & \cite{p40},\cite{p41}  \\
\hline
\end{tabular}}
\caption{\em \small Riferimenti alla letteratura primaria del secolo scorso in alcuni influenti testi di 
letteratura secondaria, che parlano dell'origine della  
quantizzazione per le particelle a spin $\nicefrac{1}{2}$. 
Parte superiore della tabella, libri di testo. Parte inferiore della tabella, resoconti storici.
La quarta colonna si riferisce alla procedura di quantizzazione dei fermioni di Majorana,
di regola ignorata nella letteratura secondaria.}\label{checz}
\end{table}

 \section{Majorana 1937 nella letteratura}\label{appb}
 
La tabella~\ref{checz} mostra che, nella letteratura secondaria riguardante i primi passi della quantizzazione canonica 
delle particelle con spin $\nicefrac{1}{2}$, 
 i contributi di Jordan,  Weisskopf e soprattutto Pauli  
vengono  riconosciuti mentre
quello di Majorana di solito no.
Questo \`e vero in particolare nella letteratura di lingua  tedesca e anche inglese, e salta agli occhi   
dalla tabella storica sinottica nella introduzione di \cite{s94}.
L'eccezione \`e il resoconto di Maurice Jacob, che fu  studente di uno dei  ragazzi di via Panisperna: Gian Carlo Wick,
fisico teorico, che capiva e rispettava Majorana. 

\`E ragionevole supporre che a causare tale curiosa sottovalutazione  abbiano contribuito il mancato riconoscimento 
da parte di  un altro ragazzo di via Panisperna, Giulio Racah \cite{g37}, che lo cita, ma non per la procedura di quantizzazione;
il suo  mentore, Wolfgang Pauli,  non cita affatto Majorana
 nel lavoro sulla connessione tra spin e statistica  \cite{p40},  e  nella notissima rassegna  del 1941 \cite{p41} gli riconosce solamente 
(e con qualche riserva) di aver  suggerito un'ipotesi originale  sui neutrini utilizzando le moderne procedure di quantizzazione,   
non di aver  proposto per primo la quantizzazione canonica 
dei fermioni.
Tuttavia, \`e evidente che il lavoro di Majorana del 1937 non riguardi solamente i neutrini:  non per niente si intitola
\begin{quote}
 {\em Teoria simmetrica dell'elettrone e del positrone.}  
\end{quote}

\section{Sulla rappresentazione di Majorana}\label{appa}
Come abbiamo argomentato nel testo, la scelta di Majorana delle matrici $4\times 4$ 
 aiuta a  
capire bene la connessione tra l'equazione di Dirac e l'esistenza degli anti-elettroni. Ma siccome questa rappresentazione 
non \`e molto utilizzata,
  \`e bene soffermarsi un po' per familiarizzarsi con essa. 
Segue qualche  osservazione utile a questo scopo, ad integrazione di quelle nelle 
appendici di~\cite{what21} e di \cite{zema}.

Iniziamo con delle considerazioni generali, partendo dalla definizione delle matrici $\gamma$
\begin{equation}
\gamma^0=\beta \mbox{ e } \vec{\gamma}= \beta\, \vec{\alpha}
\quad
\Leftrightarrow
\quad
\beta=\gamma^0\mbox{ e }\vec{\alpha}=\gamma^0\, \vec{\gamma}\end{equation} 
dove $(\vec{\gamma})^i=\gamma^i$ ($i=1,2,3$).
Tutte le scelte di matrici gamma sono equivalenti,
a patto che obbediscano alle condizioni di hermiticit\`a $(\gamma^0)^\dagger=\gamma^0$ e 
 $(\vec{\gamma})^\dagger=- \vec{\gamma}$ 
ed alle  caratteristiche propriet\`a
\begin{equation}
\{\ \gamma_\mu\, , \, \gamma_\nu \ \}=\mathbb{1}_{\tiny 4\times 4}\ 2\, g_{\mu\nu}  \mbox{ dove }g=\mbox{diag}(+1,-1,-1,-1)
\label{gm}
\end{equation}
La prima  dimostrazione di questo  fatto (a volte detto teorema di Pauli) 
si trova qui \cite{p36-1};  l'appendice di~\cite{zema} ne fornisce una 
 costruttiva, esplicita e diretta 
 ed una brevissima e di natura formale.
Le funzioni d'onda (o i campi) spinoriali realizzano
le trasformazioni di Lorentz  come segue 
\begin{equation}
\varphi'=\Lambda(\omega)\ \varphi
\end{equation}
dove
\begin{equation}
\Lambda(\omega)=\exp\left( - \frac{i}{4} \omega_{\mu\nu} \, \Sigma^{\mu\nu} \right) \mbox{ con } 
 \Sigma^{\mu\nu}=\frac{i}{2}[\, \gamma^\mu , \gamma^\nu\, ] 
\end{equation}
$\Sigma_{0i}^\dagger=- \Sigma_{0i}$ rappresentano le trasformazioni di velocit\`a  e $\Sigma_{ij}^\dagger=\Sigma_{ij}$ le rotazioni.
Per il coniugato di Dirac $\bar{\varphi}\equiv\varphi^\dagger \gamma_0$ avremo
\begin{equation}
\bar{\varphi}'= 
\bar{\varphi}\ \Lambda(-\omega)=
\bar{\varphi}\ \Lambda(\omega)^{-1}
\label{cdd}
\end{equation}


\subsection{Espressioni esplicite delle matrici gamma}

Esibiamo in tabella~\ref{tab1} varie  matrici $\gamma$ di Dirac 
che soddisfano la propriet\`a di Majorana
\begin{equation}
\hat{\gamma}_\mu^*= -\hat{\gamma}_\mu\mbox{ dove } \mu=0,1,2,3\label{mjr}
\end{equation}
Tali matrici sono scritte  come prodotti tensori (o di Kronecker) di matrici hermitiane $2\times 2$, dove 
 $\sigma_0=\mathbb{1}$ \`e la matrice identit\`a e $\sigma_{1,2,3}$ le matrici di Pauli.
\`E facile verificare le regole di anticommutazione e le propriet\`a di hermiticit\`a. 
Rimandiamo  all'appendice di~\cite{what21} per  maggior  discussione.

\begin{table}[t]
\centerline{\begin{tabular}{c|ccc}
  $\gamma^0$  & $\gamma^1$  & $\gamma^2$  & $\gamma^3$ \\ \hline
$\sigma_2\circ  \sigma_1 $ &  $ i         \sigma_2\circ  \sigma_2$   &  $i    \sigma_1\circ  \sigma_0$  &  $i \sigma_3\circ  \sigma_0$ \\
$\sigma_1\circ  \sigma_2 $ &  $ i         \sigma_1\circ  \sigma_1$ &  $i \sigma_1\circ  \sigma_3$  &  $i    \sigma_3\circ  \sigma_0$   \\
 $\sigma_2\circ  \sigma_0 $   & $ i         \sigma_1\circ  \sigma_1$ &  $i \sigma_1\circ  \sigma_3$  &  $i    \sigma_3\circ  \sigma_0$   \\
 \end{tabular}}
 \caption{\em\small Tre forme di Majorana delle matrici $\gamma$, con la stessa
 espressione di  $\gamma^3$  (che \`e diagonale), espresse come prodotti tensori delle matrici di Pauli.
In  ognuna delle forme qui indicate, possiamo
 a) cambiare il segno di qualsiasi matrice;  b)~permutare la posizione degli indici 1 e 3, oppure
 la prima e la seconda matrice di Pauli nel prodotto tensore;
 c) scambiare a piacere  tra di loro le ultime tre matrici~$\gamma$. 
 }\label{tab1}
 \end{table}

\subsection{Coniugato covariante}
In questa sezione, consideriamo solo matrici $4\times 4$.
Consideriamo una qualsiasi rappresentazione delle matrici $\hat{\gamma}$ che soddisfi la condizione di Majorana come in eq.~\ref{mjr}, 
 ed una seconda rappresentazione ${\gamma}$
ad essa collegata per mezzo di una trasformazione unitaria
\begin{equation}
{\gamma}_\mu=U\, \hat{\gamma}_\mu\, U^\dagger
\quad
\Leftrightarrow 
\quad 
{\gamma}_\mu \, U=  U\, \hat{\gamma}_\mu 
\mbox{ dove }U U^\dagger=\mathbb{1} 
\end{equation}
Evidentemente, entrambe le rappresentazioni soddisferanno la condizione caratteristica delle matrici gamma di eq.~\ref{gm}.
Le funzioni d'onda (o i campi) sono connesse come segue
\begin{equation}
{\varphi}=U\hat{\varphi}
\end{equation}
Mentre nel caso della rappresentazione di Majorana vale
$
\hat{\Lambda}^*=\hat{\Lambda}
$
questo non avviene in generale;
$
{\Lambda}^*\neq {\Lambda}
$.
Tuttavia, partendo dall'espressione 
$
{\Lambda}=U \ \hat{\Lambda}\ U^\dagger
$
verifichiamo agevolmente che
\begin{equation}
{\Lambda}^*=B^\dagger \ {\Lambda}\  B
\quad
\Leftrightarrow 
\quad 
B\ {\Lambda}^*= {\Lambda}\  B
\end{equation}
dove
\begin{equation}
B\equiv U U^t\label{defa} 
\end{equation}
 una matrice che \`e sia unitaria che simmetrica. 
Pertanto, definendo lo {\em spinore coniugato,}
constatiamo che esso si trasforma esattamente come lo spinore ${\varphi}$:
\begin{equation}
{\varphi}^c\equiv B\ {\varphi}^*
\qquad \Rightarrow\qquad
({\varphi}^c)'={\Lambda}\ {\varphi}^c
\end{equation}
In modo simile, da 
$U^\dagger\, {\gamma}_\mu\, U=\hat{\gamma}_\mu=-\hat{\gamma}_\mu^*=-U^t\, {\gamma}_\mu^*\, U^*$, desumiamo
\begin{equation}
{\gamma}^*_\mu=-B^\dagger\ {\gamma}_\mu\  B
\quad
\Leftrightarrow 
\quad 
B\ {\gamma}^*_\mu=- {\gamma}_\mu\  B \label{miamo}
\end{equation}

\subsection{Coniugazione di carica}
Introduciamo  la matrice (unitaria) di coniugazione di carica $C$ come segue
\begin{equation}
{\varphi}^c= C\   \overline{ {\varphi}}^t
\end{equation}
Dalla definizione del coniugato di Dirac, eq.~\ref{cdd}, ed usando la definizione \ref{defa} abbiamo:
\begin{equation}
C= B\ {\gamma}_0^*  \label{defc}
\end{equation}
Inoltre, prendendo 
il complesso coniugato della relazione
$U^\dagger {\gamma}_0= \gamma_0\, U^\dagger$,  troviamo che:
\begin{equation}
C= U {\gamma}_0^* U^t
\end{equation}
Da questa espressione e dalla propriet\`a delle matrici di Majorana, eq.~\ref{mjr},  
segue subito che 
\begin{equation}
C^t=-C
\end{equation}
 Da ${\gamma}_\mu^\dagger={\gamma}_0^\dagger\, {\gamma}_\mu\, {\gamma}_0$
che implica ${\gamma}_\mu^t={\gamma}_0^t \,{\gamma}_\mu^*\, {\gamma}_0^*$,  assieme 
alle eqq.~\ref{miamo} e \ref{defc}, deduciamo poi che, 
\begin{equation}
{\gamma}^t_\mu=-C^\dagger\ {\gamma}_\mu\  C
\quad
\Leftrightarrow 
\quad 
C\ {\gamma}^t_\mu=- {\gamma}_\mu\  C \label{miamo2}
\end{equation}
Le ultime due propriet\`a  ricorrono frequentemente 
nel contesto della fisica che estende il modello standard delle interazioni elettrodeboli. 
Simili relazioni valgono per le matrici di trasformazione ${\Lambda}$.

\footnotesize

\tableofcontents
\end{document}